\documentclass[12pt]{article}

\usepackage{amsfonts}
\usepackage{epsfig}
\usepackage{latexsym}
\usepackage{graphicx}
\def\bequ{\begin{equation}}
\def\eequ{\end{equation}}
\def\barr{\begin{array}}
\def\earr{\end{array}}
\def\half{{1\over 2}}
\def\ben{\begin{equation}}
\def\een{\end{equation}}
\def\bena{\begin{eqnarray}}
\def\eena{\end{eqnarray}}

\newcommand{\sect}[1]{\setcounter{equation}{0}\section{#1}}

\def\b1{e^0}

\newcommand{\be}{\begin{equation}}
\newcommand{\ee}{\end{equation}}
\def\bea{\begin{eqnarray}}
\def\eea{\end{eqnarray}}

\def\del {\partial}
\def\nn{\nonumber}

\def\half {{1 \over 2}}

\def\lesssim{\mathrel{\hbox{\rlap{\hbox{\lower4pt\hbox{$\sim$}}}\hbox{$<$}}}}
\def\gtrsim{\mathrel{\hbox{\rlap{\hbox{\lower4pt\hbox{$\sim$}}}\hbox{$>$}}}}
\def\calF {{\cal F}}

\def\Pf{\mbox{Pf}}
\def\Ful{\underline{F}}

\def\thetaul{\underline{\theta}}

\def\u0{{\underline 0}}

\def\url{{\underline {r+\ell}}}

\begin{document}

\begin{flushright}
SU-ITP/02-06\\
\hfill{ hep-th/0203019}\\
\today
\end{flushright}

\vspace{1cm}

\begin{center}

{\Large \bf    D3/D7 Inflationary Model and  M-theory}

\vspace{2cm}

{\bf Keshav Dasgupta, Carlos Herdeiro, Shinji Hirano and Renata
Kallosh}\footnote{Email addresses: keshav,carlos,hirano@itp.stanford.edu; kallosh@stanford.edu}
\vspace{.5cm}

{\it Department of Physics}

{\it Varian Laboratory of Physics}

{\it Stanford University, Stanford, California 94305, USA}

\vspace{2cm}

\underline{ABSTRACT}

\begin{quote}

A proposal is made for a cosmological  D3/D7 model with a constant
magnetic flux along the D7 world-volume. It describes an
$\mathcal{N}=2$ gauge model with Fayet-Iliopoulos terms and the
potential of the hybrid P-term inflation. The motion of the
D3-brane towards D7 in a phase with  spontaneously broken
supersymmetry   provides a period of slow-roll inflation in the de
Sitter valley, the role of the inflaton being played by the
distance between  D3 and D7-branes. After tachyon condensation a
supersymmetric ground state is formed: a D3/D7 bound  state
corresponding to an Abelian non-linear  (non-commutative)
instanton. In this model the existence of a non-vanishing
cosmological constant is associated with the resolution of the
instanton singularity.  We discuss a possible embedding of this
model into a compactified M-theory setup.

\normalsize

\end{quote}

\end{center}

\newpage
\sect{Introduction}

 Hybrid inflation \cite{Linde:1991km} can be
naturally implemented in the context of supersymmetric theories
\cite{Copeland:1994vg}-\cite{Linde:1997sj}. The basic feature of
these inflationary models is the existence of two phases in the
evolution of the universe: a slow-roll inflation in the de Sitter
valley of the potential (the Coulomb phase of the gauge theory)
and a  tachyon condensation phase, or `waterfall stage', towards
the ground state Minkowski vacuum (a Higgs phase in gauge theory).

In $\mathcal{N}=1$ supersymmetric theories, hybrid inflation may
arise as F-term inflation \cite{Copeland:1994vg,Dvali:1994ms} or
D-term inflation \cite{Binetruy:1996xj}. In $\mathcal{N}=2$
supersymmetric theories there is a triplet of Fayet-Iliopoulos
(FI) terms, $\xi^r$, where $r=1,2,3$. Choosing the orientation of
the triplet of FI terms, $\xi^r$, in directions 1,2, F-term
inflation is promoted to $\mathcal{N}=2$ supersymmetry
\cite{Watari:2000jh}. The more general case with ${\mathcal N}=2$,
when all 3 components of the FI terms are present, is called
\textit{P-term inflation} \cite{Kallosh:2001tm}. When  $\xi^3$ is
non-vanishing, a special case of D-term inflation with Yukawa
coupling related to gauge coupling is recovered. In this fashion,
the two supersymmetric formulations of hybrid inflation are
unified in the framework of ${\mathcal N}=2$ P-term inflation.
This gauge theory has the potential \cite{Kallosh:2001tm} \be V=
{g^2\over 2}\Phi^\dagger \Phi (A^2+B^2) - \left[{1\over 2} (P^r)^2
+P^r\left({g\over 2} \Phi^\dagger \sigma^r
\Phi+\xi^r\right)\right] \ ,\ee where $P^r$ is a triplet of
auxiliary fields of the $\mathcal{N}=2$ vector multiplet,
$\Phi^\dagger, \Phi$ are 2 complex scalars forming a charged
hypermultiplet, $A, B$ are scalars from the $\mathcal{N}=2$ vector
multiplet and $g$ is the gauge coupling. The auxiliary field
satisfies the equation $P^r= - (g\Phi^\dagger \sigma^r
\Phi/2+\xi^r)$ and the potential simplifies to \be V= {g^2\over
2}\left[ \Phi^\dagger \Phi (A^2+B^2) +\left({1\over 2}
\Phi^\dagger \sigma^r \Phi+{\xi^r\over g}\right)^2\right] \ . \ee

An additional advantage of using $\mathcal{N}=2$ supersymmetric
models  for inflation is the possibility to link it to M/string
theory where cases with $\mathcal{N}=2$ supersymmetry are simpler
and less arbitrary than the cases with $\mathcal{N}=1$
supersymmetry.

In our first attempt\footnote{Earlier studies of brane inflation
were performed in  \cite{dvaliedi,BAB}.} to link string theory to
a gauge model with a hybrid potential \cite{Herdeiro:2001zb}, we
used a system with a D4-brane attached to NS5-branes and having a
small angle relative
 to a D6-brane, so that the Coulomb phase of the theory is slightly
non-supersymmetric and forces the D4 to move towards the D6-brane
(see also \cite{Kyae:2001mk}). This setup reproduces accurately
the properties of the non-supersymmetric de Sitter vacuum  of
P-term inflation, for which $(P^r)_{\rm deSit}= -\xi^r$ and
$(V)_{\rm deSit}= \vec\xi^2/ 2$. In particular, the mass splitting
of the scalars in the hypermultiplet (e.g. for the case of $\xi^3=
\xi\neq 0$) \be M^2_{\rm hyper}= g^2(A^2+B^2)\pm g{\xi}
\label{split}\ , \ee is reproduced by the low-lying string states;
the attractive force between the D4 and D6-brane is a one loop
effect from the open string channel, and correctly reproduces the
one-loop gauge theory potential \be \Delta V= {\xi^2 g^2\over
16\pi^2} \ln {|A^2+B^2|\over |A^2+B^2|_c} \ , \label{1loop}\ee for
large values of the inflaton field. Notice the inflaton is the
distance between D4 and D6 in the brane model.

When the distance between the branes becomes smaller than the
critical distance \be |A^2+B^2|_c={\xi\over g} \label{bifurc} \
,\ee the spectrum of 4-6 strings develops a tachyon. The tachyon
condensation is associated to a phase transition. A final Higgs
phase with unbroken $\mathcal{N}=2$ supersymmetry is described in
this model by a reconfiguration of branes: D6 cuts D4 into two
disconnected parts, so that $\mathcal{N}=2$ supersymmetry is restored. From the
viewpoint of the gauge theory it is described by  a vanishing
auxiliary field $ (P^r)_{\rm susy}= 0$, a vanishing vev of the
inflaton field $A=B=0$,  and a vanishing potential $(V)_{\rm
susy}= 0$. The hypermultiplet vev is not vanishing and has to
satisfy  \be {g\over 2} \Phi^\dagger \sigma^r \Phi=- \xi^r \ .
\label{ADHM}\ee

One attractive feature of D4/D6/NS5 model is that the deviation
from supersymmetry in the Coulomb stage can be very small and
supersymmetry is spontaneously broken. Nevertheless, a large
number of e-foldings can be produced within a D-brane.  Given the
lessons from the D-brane approach to black hole physics, such
small deviation from supersymmetry might be an advantage over
brane/anti-brane models \cite{BAB}, for which the deviation from
supersymmetry is large.

The main difference between our model \cite{Herdeiro:2001zb} and
other models of brane inflation
\cite{Kyae:2001mk,dvaliedi,BAB,Garcia-Bellido:2001ky,Burgess:2001vr}
is that our model provides the  brane description of the full
potential of hybrid P-term inflation. This includes both the
logarithmic quantum corrections to the Coulomb branch potential
and the exit from inflation with the corresponding supersymmetric
Minkowski vacuum.

The first purpose of this paper is to suggest a new model
(partially dual to \cite{Herdeiro:2001zb}) which gives a better
description of the Higgs branch (exit from inflation) keeping the
nice properties of the Coulomb branch (inflation). We will then
argue that the stringy understanding of the exit from inflation
provides a possible explanation for the existence of a
non-vanishing cosmological constant. The second purpose is to
suggest how this new model may be consistently compactified to
four dimensions, without spoiling its cosmological properties.

The model consists of a D3-brane parallel to a D7-brane at some
distance, which again is the inflaton field. The supersymmetry
breaking parameter is related to the presence of the antisymmetric
$\mathcal{F}_{mn}$ field on the worldvolume of the D7-brane, but
transverse to the D3-brane. When this field is not self-dual in
this four dimensional space, the supersymmetry of the combined
system is broken. This is (to some extent) a type IIB dual version
of the cosmological model proposed in \cite{Herdeiro:2001zb},
which guarantees that the good properties in the Coulomb stage are
kept; in particular the spectrum and the attractive potential
should match the ones of P-term inflation. One immediate
simplification is that the NS5-branes are not needed any longer.

More interestingly is the understanding of the Higgs stage.
Equation (\ref{ADHM}) can be associated with the
Atiyah-Drinfeld-Hitchin-Manin construction of instantons with
gauge group $U(N)$. The moduli space of one instanton is the
moduli space of vacua of a $U(1)$ gauge theory with $N$
hypermultiplets and (\ref{ADHM}) is the corresponding ADHM
equation. This suggests that in the brane construction  of the
cosmological model we may look for some instantons on the
worldvolume of the brane in the Higgs phase of the theory.  We
will find an abelian non-linear instanton solution with associated
ADHM equation (\ref{ADHM}). Moreover, the presence of a
cosmological constant will translate as the resolution of the
small size instanton singularity.

One other nice feature of the D3/D7 cosmological model is that it
is well explained in terms of $\kappa$-symmetry of the D7-brane
action, both in Coulomb phase as well as in the Higgs phase.

Unbroken supersymmetry of bosonic configurations in supergravity
has already proved to be an important tool in our understanding of
M/string theory. To  obtain BPS solutions of supergravity one has
to find Killing spinors, satisfying a condition $\delta \psi =0$
for all fermions and find out how many non-vanishing spinors solve
this equation for a given solution of bosonic equations. This
allows to find  configurations with  some fraction of unbroken
supersymmetry in supergravity.

Unbroken supersymmetry of the bosonic configurations on the
worldvolume of the $\kappa$-symmetric branes embedded into some
curved space (on shell superspace) was studied less so far. The
condition of supersymmetric vacua in the context of the
supermembrane  was suggested in \cite{Bergshoeff:1987dh}.  This
condition was introduced more recently in the context of
supersymmetric cycles in \cite{Becker:1995kb}. Later on, when more
general $\kappa$-symmetric brane actions were discovered in
\cite{Cederwall:1996pv}, it was established in
\cite{Bergshoeff:1997kr} that for all cases of D-branes, M2 and M5
$\kappa$-symmetric branes, a universal equation for the BPS
configurations on the worldvolume  is given by $
(1-\Gamma)\,\epsilon =0. \label{susy} $ Here $\Gamma(X^\mu
(\sigma),  \theta(\sigma), A_i(\sigma))$ is a generator of
$\kappa$-symmetry and it should be introduced into the equation
for unbroken supersymmetry with vanishing value of the fermionic
worldvolume field $\theta(\sigma)$. The existence or absence of
solutions to these equations will fit naturally in the two stages
of our cosmological model.

In addressing compactification of such cosmological model to four
dimensions, as to recover four dimensional gravity, one is
inevitably faced with the issue of anomalies. In many cases this
is associated to the requirement that the overall charge in a
compact space must vanish. We will point out that we can face our
D3/D7 model within a more general setup, related to F-theory
\cite{Vafa:1996xn, sen96,bds}, where the D7-brane charge is
cancelled by orientifold 7-planes or $(p,q)$ 7-branes. This seems
to provide a setup where, in string theory, the compactification
could be consistently performed\footnote{Some recent papers which
also consider orientifold planes to describe the inflationary
scenario are \cite{Burgess:2001vr}.}. This is a particularly
relevant issue for the cosmological models with FI terms. It was
shown by Freedman \cite{Freedman:1976uk} that FI terms in
$\mathcal{N}=1$, $D=4$ supergravity must be accompanied by an
axial gauging of gravitino and  gaugino which makes the theory
anomalous. We then discuss the uplifting of the compactified,
anomaly free IIB model to M-theory, where it becomes simpler. We
close with a discussion.

\sect{De Sitter valley: separated D3/D7 system with fluxes}

We will start by reviewing the spectrum of the D3/D7 system with a
gauge field \cite{Berkooz:1996km,Seiberg:1999vs}. The boundary
conditions on the D7 side will depend on a gauge field on its
surface. To understand the dynamics of the branes we will consider
a D7-brane in a background of the D3-brane. We will allow a
worldvolume gauge field on D7 and find that  supersymmetry is
broken when this gauge field is not self-dual.  The potential
between branes will be shown to have a logarithmic dependence on
the distance, as expected for broken supersymmetry.

In an alternative picture, a D3-brane will probe  the background
of a D7-brane with a  bulk $B$ field. The combined system will be
supersymmetric  under condition that the $B$ field is self-dual.
For the non-self-dual field we calculate the potential and find
again a logarithmic dependence on the distance. Therefore we
recover from the probe approximation, the gauge theory and open
string theory results \cite{Herdeiro:2001zb}.

\subsection{Perturbative String theory analysis}

Consider a type IIB system with $D3$ and $D7$-branes plus a
constant worldvolume gauge field ${\cal F}$ field along directions
\begin{center}
 \begin{tabular}{*{11}{|c}|}
 \hline
 \mbox{ } & 0 & 1 & 2 & 3 & 4 & 5 & 6 & 7 & 8 & 9 \\ \hline
 $D3$ & $\times$ & $\times$ & $\times$ & $\times$ & $\hspace{0.3cm}$ & $\hspace{0.3cm}$ & \mbox{} & \mbox{} & \mbox{} &  \\ \hline
 $D7$ & $\times$ & $\times$ & $\times$ & $\times$ & \mbox{  } & \mbox{  } & $\times$ & $\times$ & $\times$ & $\times$ \\
 \hline ${\cal F}$ & \mbox{  } & \mbox{  } & \mbox{  } & \mbox{  } & \mbox{  } & \mbox{  }
 &  $\times$ & $\times$ & $\times$ & $\times$  \\ \hline
 \end{tabular}
\end{center}
\begin{center}
Table 1
\end{center}
This is illustrated in figure 1. It describes a de Sitter stage of
a hybrid inflation and it is T-dual to the type IIA $D4/D6$ model
of branes at an angle (without the NS5-branes), considered in
\cite{Herdeiro:2001zb}.

\begin{figure}[h!]
\begin{picture}(75,0)(0,0)
\put(100,135){${\mathcal F}$}  \put(105,60){$D7$}
\put(130,0){$x^{4,5}$} \put(220,60){$D3$}
 \put(45,30){$x^{0,1,2,3}$} \put(115,25){$x^{6,7,8,9}$}
 \put(100,90){$_{\sigma=\pi}$} \put(245,112){$_{\sigma=0}$}
\end{picture}
\centering \epsfysize=13cm
\includegraphics[scale=0.5]{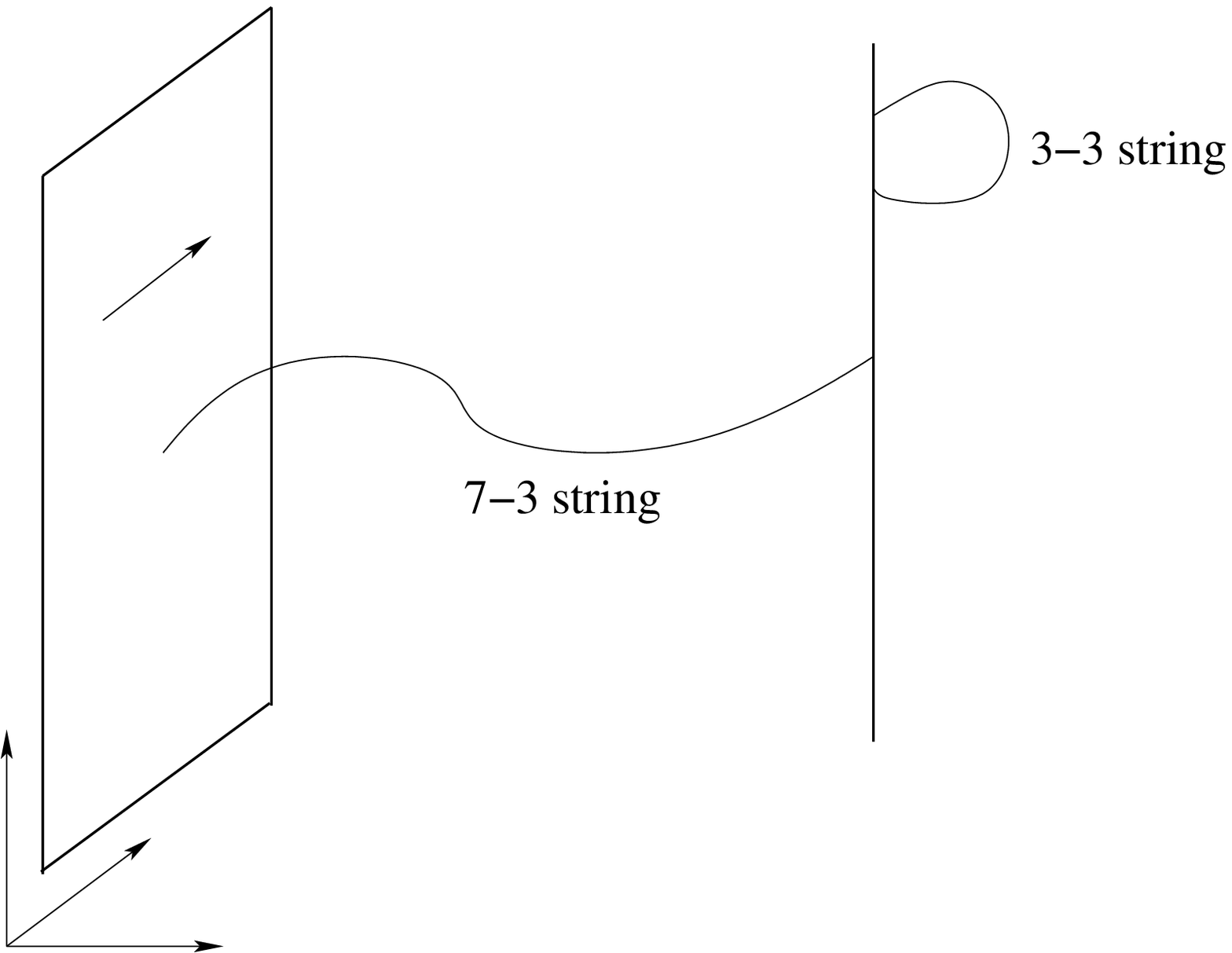}
\caption{The D3/D7 ``cosmological" system. The 3-3 strings give
rise to the ${\mathcal N}=2$ vector multiplet, the 7-3 strings to
the hypermultiplet and the worldvolume gauge field ${\mathcal F}$
to the FI terms of the $D=4$ gauge theory.}
\end{figure}

We place D7  at $ (x^4)^2+(x^5)^2=0$ and D3 is initially  at some
$d^2= (x^4)^2+(x^5)^2 >> d_c^2$, where $d_c$ is defined below in
(\ref{dcritical}). There is a constant worldvolume gauge field
${\cal F}=dA-B$ present on D7:
\begin{equation}
 {\cal F}_{67}=\tan{\theta_1},~~~~{\cal F}_{89} = \tan{\theta_2} \
 , \label{fwv}
\end{equation}
responsible for spontaneous breaking of supersymmetry. For example
we may have $B=0$ in the bulk and the following vector fields on
the brane: \be A_6=-{1\over 2}\tan{\theta_1} x^7 \ , \quad
A_7={1\over 2}\tan{\theta_1} x^6 \ , \quad A_8=-{1\over
2}\tan{\theta_2} x^9 \ , \quad A_9={1\over 2}\tan{\theta_2} x^8 \
. \label {magnetic}\ee Note that if ${\cal F}$ is self-dual,
supersymmetry would be unbroken \cite{Berkooz:1996km}. This will
be explained via $\kappa$-symmetry in the next subsection. For
future reference we define \bequ {\mathcal F}^{\pm}={\mathcal
F}\pm\star{\mathcal F} \ , \eequ and similarly for $B$.

Let us define complex coordinates:
\begin{equation}
z^1 = x^6+ix^7, ~~~~~z^2= x^8+ix^9 \ .
\end{equation}
The boundary conditions on the D3/D7 strings along the $6,7,8,9,$
directions are therefore:
\begin{equation}
(\del_{\tau}z^k)_{\sigma=0}=0, ~~~~~~
(\del_{\sigma}z^k+\tan{\theta_k} \del_{\tau}z^k)_{\sigma=\pi}=0 \
,
\end{equation}
(no sum on $k$) where $k=1,2$, and similar conditions for ${\bar
z^k}$ (with $\tan{\theta_k} \to -\tan{\theta_k}$). Let us now
write the mode expansion as:
\begin{equation}
z^k=\sum_n A_{n+\nu_k}^k e^{(n+\nu_k)(\tau+i\sigma)}+ \sum_n
B_{n+\nu_k}^ke^{(n+\nu_k)(\tau-i\sigma)} \ .
\end{equation}
The first boundary condition yields
\begin{equation}
A_{n+\nu_k}^k= - B_{n+\nu_k}^k \ ,
\end{equation}
while the second one gives
\begin{equation}
e^{2\pi i \nu_k} = -{1+i\tan{\theta_k}\over 1-i\tan{\theta_k}} \ \
\ \Leftarrow \ \ \nu_k = {1\over \pi} \left(\theta_k +
\frac{\pi}{2}\right) \ .
\end{equation}
We take $-\pi/2<\theta_k<\pi/2$, which implies that $0<\nu_k<1$.
It only remains to find out the zero-point energy as a function of
$\nu_k$, in the NS sector. We use:
\begin{equation}
\sum_{n=1}^{\infty} (n- \nu) = -{1\over 12}(6\nu^2 - 6\nu +1) \ .
\end{equation}
The bosons along directions $x^{6,7}$ and $x^{8,9}$ are quantized
with mode numbers $n\pm\nu_k$ and the fermions have mode numbers
$n \pm (\nu_k - {1 / 2} )$. Thus the zero point energy of the
system will be
\begin{equation}
E = -{1\over 2} {\Big(} |\nu_1 - {1\over 2}| + |\nu_2 - {1\over
2}| {\Big)} \ .
\end{equation}

\noindent From the above equation we have four possible cases:

\begin{center}
 \begin{tabular}{*{9}{|c}|}
 \hline
 $\mbox{}$ & $\nu_1$ & $\nu_2$ &  $E$ \\
 \hline
1. &$\nu_1 > 1/2$ & $\nu_2 < 1/2$ & $-(\nu_1 - \nu_2)/2$   \\
\hline
2. & $\nu_1 < 1/2$ & $\nu_2 > 1/2$  & $(\nu_1 - \nu_2)/2$ \\
 \hline
3. & $\nu_1 > 1/2$ &  $\nu_2 > 1/2$  &  $-(\nu_1+\nu_2-1)/2$ \\
\hline 4. & $\nu_1 < 1/2$ & $\nu_2 < 1/2$ & $(\nu_1+\nu_2-1)/2$
\\ \hline
 \end{tabular}
\end{center}
\begin{center}
Table 2 \end{center}
 \noindent We can choose the GSO even states to be either 1,2 or 3,4. We will choose the former.
 Then the zero point energy becomes
\begin{equation}
E = \pm {1\over 2}(\nu_1-\nu_2) = \pm {\theta_1-\theta_2\over
2\pi} \ .
\end{equation}
Therefore the lowest lying multiplet of states of open strings consist of
bosons whose masses are given by
\begin{equation}
M^2_{\pm}= {d^2\over (\pi \alpha')^2}\pm {\theta_1-\theta_2\over
2\pi \alpha'} \ ,
\end{equation}
where we observe that the boson of mass $M^2_{-}$ becomes
tachyonic at the critical distance\footnote{We are assuming
$\theta_1 > \theta_2$.}
\begin{equation}
d^2_c \equiv \pi \alpha'{\Big(}{\theta_1-\theta_2 \over 2}{\Big) \
, } \label{dcritical}
\end{equation}
and the other boson remains at positive mass. The fermion in the multiplet
have a mass
\begin{equation}
M_{\psi}^2= {d^2\over (\pi \alpha')^2} \ ,
\end{equation}
as the contribution from the zero point energy to the Ramond
sector is zero.

The higher states in the string spectrum satisfying the condition
of spontaneously broken supersymmetry $Str M^2=0$, also have
tachyonic masses whose values and splitting are given in
{\cite{Herdeiro:2001zb}. One can check however that it is the very
first tachyon which provides a critical distance between the
branes, all other higher level tachyons always come at distances
smaller than the first one. Therefore already at $d^2_c$ there is
a phase transition and a waterfall stage sets on as the potential
of the gauge theory suggests.

The ${\cal F}$ field plays the role of the Fayet-Illiopoulos term,
from the viewpoint of the field theory living on the $D3$-brane.
It creates an instability in the system driving the $D3$-brane
into the $D7$-brane; this is the de Sitter stage. The evolution
follows the description given in \cite{Herdeiro:2001zb}. In
particular, a tachyon will form and the system will end in a
supersymmetric configuration which will be analyzed in section 3.

\subsection{D7-brane worldvolume analysis and $\kappa$-symmetry}
In the presence of Fayet-Illiopoulos parameter, the system of $D3$
and $D7$-branes at some distance from each other is unstable. It
is instructive to see this instability from the viewpoint of the
D7-brane worldvolume action

\begin{equation} \mathcal{S}=-T_7\int d^8\sigma
e^{-\phi}\sqrt{-\det{(g+\mathcal{F})}}+T_7\int \sum
\mathcal{C}_{p+1}\wedge e^{\mathcal{F}}, \label{D7}
\end{equation} where $g$ is the metric induced on the brane and
${\cal F}= F-B$, where $B$ is a pull-back of the NS-NS 2-form and
$F$ is the 2-form ($F=dA$) Born-Infeld field strength. We place
our $D7$-brane in the
 $D3$-brane background. The dilaton is constant while the
metric and self-dual RR form read
\begin{equation}
\begin{array}{c}
\displaystyle{ds^2=H^{-1/2}ds^2({\mathbb
E}^{3,1})+H^{1/2}ds^2({\mathbb E}^6)}, \\\\
\displaystyle{{\mathcal{F}}^{RR}=\partial_{i}H^{-1} dx^i\wedge
\epsilon_{{\mathbb {E}^{3,1}}}+ \star(\partial_{i}H^{-1}
dx^i\wedge \epsilon_{{\mathbb{E}}^{3,1}})}.
\end{array}
\end{equation}
$H$ is the usual single center harmonic function on ${\mathbb
E}^6$, $H=1+Q/r^4$, and $\epsilon_{{\mathbb{E}}^{3,1}}$ is the
volume form on ${\mathbb E}^{3,1}$. The embedding functions for
the $D7$-brane are $x^{\mu}=\sigma^{\mu}$, where
$\mu=0,1,2,3,6,7,8,9$. We also turn on the worldvolume gauge
field, with components (\ref{fwv}). Denoting by ${\mathcal{V}}_3$
the volume along $\sigma^1,\sigma^2,\sigma^3$ and allowing
$i=6,7,8,9$, we find the effective potential,
\begin{equation}
V=T_{7} {\mathcal{V}}_3\int
d\sigma^{i}\left[\sqrt{(1+H^{-1}\tan{\theta_1}^2)(1+H^{-1}\tan{\theta_2}^2)}-(H^{-1}-1)\tan{\theta_1}
\tan{\theta_2}\right]. \end{equation}

Clearly, if $\theta_1=\theta_2$, the force between the $D3$ and
$D7$ vanishes. This corresponds to a self dual configuration in
the Euclidean four space $\mathbb{R}^4$ parameterized by
directions 6 to 9. Let $\rho$ be the radial coordinate in this
four space. Also denote $ d^2\equiv(x^4)^2+(x^5)^2 $. Change
coordinates to $\lambda=\rho^2+d^2$. For large distances,
$d^4>>Q$, and up to a divergent term coming from the infinite
volume of the brane, the potential can be rewritten as \bequ
\barr{l}
\displaystyle{V\simeq-\frac{\pi^2}{2}QT_7{\mathcal{V}}_3\left(\sin{\theta_1}-\sin{\theta_2}\right)^2\int^{\Lambda}_{d^2}\frac{d\lambda}{\lambda}}
\ , \\\\
\ \ \ \
\displaystyle{=\frac{\pi^2}{2}QT_7{\mathcal{V}}_3\left(\sin{\theta_1}-\sin{\theta_2}\right)^2\ln{\frac{d^2}{\Lambda}}
} \ , \earr \eequ where we introduced the cutoff $\Lambda$. This
reproduces the attractive potential of hybrid inflation in
accordance with the results discussed in \cite{Herdeiro:2001zb}.
Note that the choice of gauge for the supergravity potential
${\mathcal{C}}_4$ is such that it vanishes at infinity and
reproduces the right energy for the D3/D7 bound state.

The breaking of supersymmetry manifest in the non-vanishing force
between D3 and D7 can be deduced from general considerations of
$\kappa$-symmetric D7-brane in the background of D3-brane. The
bosonic part of the action is given in eq. (\ref{D7}). Solutions
of the equations following from this action may have some unbroken
supersymmetry\footnote{We are using notation of
\cite{Bergshoeff:1997kr}}  if there are non-trivial spinor
solutions to \bea (1-\Gamma)\epsilon=0 \ ,\label{susywv} \eea
where the $\kappa$-symmetry projector $\Gamma$ for a D7-brane in a
D3-brane background is given by \cite{Bergshoeff:1997kr} \bea
\Gamma= e^{-{a\over 2}}\, i \sigma_2 \otimes \Gamma_{01236789}\,
e^{{a\over 2}}= e^{-{a }}\, i \sigma_2 \otimes \Gamma_{01236789}\,
\label{kappa}\ . \eea Here $\epsilon^{\alpha I}$ is a spinor of
type IIB theory which has  32 independent components. It is
represented by two  Majorana spinors $I=1,2$ and $\alpha=1, \dots,
32$, satisfying  a chirality constraint;  $(\sigma_2)_I{}^J$ is a
Pauli matrix and in the absence of the `rotation' factor $a$ this
Killing equation would just reproduce the D7-brane projector,
$(1-i \sigma_2 \otimes \Gamma_{01236789})\epsilon=0$, which would
correspond to 1/2 of unbroken supersymmetry. We are looking for a
configuration with an $F=dA$ on the worldvolume which is
skew-diagonal, as in (\ref{fwv}). The general relation between
${\cal F}$ and $a$ is rather complicated but it turns out to be
very simple in our case since the choice of the skew-diagonal
basis is possible for the matrix ${\cal F}$: it is antisymmetric
and independent on the worldvolume coordinates, \bea a=  \sigma_3
\otimes (\theta_1\gamma_{67}+ \theta_2\gamma_{89})=  \sigma_3
\otimes H^{1/2} (\theta_1 \, \Gamma_{67}+ \theta_2\,\Gamma_{89}) \
, \eea where $\gamma_i= E_i{}^a \Gamma_a$, and the vielbeins are
given by the metric of the D3-brane. $\Gamma_a$ are space-time
gamma-matrices. The presence of the ${\cal F}$ field affects the
unbroken supersymmetry projector of the D7-brane in  D3 background
and we find for the Killing spinors a condition: \bea
\exp\{-\sigma_3 \otimes H^{1/2} (\theta_1 \, \Gamma_{67}+
\theta_2\,\Gamma_{89})\} i \sigma_2 \otimes \Gamma_{01236789} \,
\epsilon =\epsilon \label{killing}\ . \eea Due to the presence of
the D3-brane background, the Killing spinor has also to satisfy
the following conditions \bea \epsilon= H^{-{1/4}}\epsilon_0 \ ,
\qquad  i \sigma_2 \otimes \Gamma_{0123} \, \epsilon_0 =\epsilon_0
\ ,\qquad \  \eea where $\epsilon_0$ is a constant spinor which
breaks half of the supersymmetry. Taking this into account in eq.
(\ref{killing}) we reduce the problem to the following one \bea
\exp \left\{ -{1\over 2} \sigma_3 \otimes H^{1/2}
\Gamma_{67}[(\theta_1 + \theta_2) (1-\Gamma_{6789})+ (\theta_1 -
\theta_2) (1+\Gamma_{6789})]\right\} \otimes \Gamma_{6789} \,
\epsilon_0 =\epsilon_0 \label{killing1} \ .\eea

The harmonic function of the D3 background at the position of the
D7 brane $ d^2=(x^4)^2+(x^5)^2=0$ has the form,  $H=1+ Q/\rho^4$
where $\rho^2=
(\sigma^6)^2+(\sigma^7)^2+(\sigma^8)^2+(\sigma^9)^2$. Equation
(\ref{killing1}) shows that unless $\theta_1 = \theta_2$ there is
no solution and all supersymmetries are broken. Indeed if
$\theta_1 = \theta_2=\theta$ we find \bea \exp\{-\sigma_3 \otimes
H^{1/2} \theta \, \Gamma_{67}(1- \Gamma_{6789})\}   \otimes
\Gamma_{6789} \, \epsilon_0 =\epsilon_0 \label{killing2}\ . \eea
The solution is $ \epsilon_0 =  \Gamma_{6789} \, \epsilon_0 . $
Thus for self-dual ${\cal F}$, ${\cal F}^-=0$  our exact
non-linear Killing condition can be satisfied for the Killing
spinors, subject to 2 projector equations \bea
&&\epsilon = i \sigma_2 \otimes \Gamma_{01236789}\epsilon \nonumber\\
&&\epsilon =    \Gamma_{6789}\epsilon \ .\eea However, if
$\theta_1 \neq \theta_2$, equation (\ref{killing1}) can not be
satisfied for non-vanishing $\epsilon_0$ : the constraint on
constant spinors can not depend on a function of the worldvolume
coordinates $H(\sigma)$.

 It is important to stress here that $H(\sigma)$ is a function of the worldvolume coordinates which is fixed by the properties of
the background. It will be quite different in the Higgs phase when
an analogous situation will occur and it will be possible to use
the independence of the corresponding function on the worldvolume
coordinates as a condition for the non-linear instanton equation.

\subsection{ D3-brane worldvolume analysis}
We may see the instability in the D3/D7 system still in a
different fashion. We take a D7-brane geometry with a B-field
obeying $B^-\neq 0$, probed by a D3-brane. A corresponding
supergravity solution  can be taken in the following form
(obtained by two T-dualities at an angle from the D5-brane
solution):

\begin{eqnarray}
ds_{D7}^2&=&Z_7^{-1/2}ds^2({\mathbb E}^{3,1})
+Z_7^{1/2}ds^2({\mathbb E}^{2}_{45})+Z_7^{-1/2}H_1ds^2({\mathbb
E}^{2}_{67}) +Z_7^{-1/2}H_2ds^2({\mathbb
E}^{2}_{89}), \nn\\
e^{2\phi}&=&g_s^2Z_7^{-2}H_1 H_2, \\
B_{67}&=&-\tan\theta_1 Z_7^{-1} H_1,\quad
B_{89}=-\tan\theta_2 Z_7^{-1}H_2,\nn\\
C_4&=&\left(Z_7^{-1}-1\right)\sin\theta_1\sin\theta_2
\epsilon_{{\mathbb {E}^{3,1}}}, \nn\\
C_6&=&\left(Z_7^{-1}-1\right)\left[H_1\cos\theta_1\sin\theta_2
     dx_6\wedge dx_7+H_2\cos\theta_2\sin\theta_1
     dx_8\wedge dx_9\right]\wedge\epsilon_{{\mathbb {E}^{3,1}}}, \nn\\
C_8&=&\left(Z_7^{-1}-1\right)H_1H_2\cos\theta_1\cos\theta_2
\epsilon_{{\mathbb {E}^{3,1}}}\wedge
     dx_6\wedge dx_7\wedge dx_8\wedge dx_9, \nn
\end{eqnarray}
where $ds^2({\mathbb E}^{2}_{ij})$ denotes the Euclidean two space
parameterized by cartesian coordinates $x^i,x^j$ and we have
defined
\begin{eqnarray}
H_1&=&\left(\cos^2\theta_1+\sin^2\theta_1 Z_7^{-1}\right)^{-1},\quad
H_2=\left(\cos^2\theta_2+\sin^2\theta_2 Z_7^{-1}\right)^{-1},\nn\\
Z_7&=&\lim_{\epsilon\to 0^+}\left(1+c_7\frac{\Gamma(\epsilon/2)}
{r^{\epsilon}}\right)=1-2c_7\ln\left(r/e^{1/\epsilon}\right).
\label{apx}
\end{eqnarray}

The $D3$-brane probe action is given by
\begin{eqnarray}
S_{D3}=-T_{D3}\int d^4\sigma e^{-(\phi-\phi_0)}
\sqrt{-\det\left(g+\calF\right)}
+T_{D3}\int C_4,
\end{eqnarray}
where the zero mode of the dilaton $\phi_0$ is defined by $e^{\phi_0}=g_s$.

For our application the gauge invariant 2-form $\calF$ is
vanishing, and we do the approximation of $Z_7^{-1}\sim
1+2c_7\ln\left(r/\Lambda\right)$ with $\Lambda=e^{1/\epsilon}$ for
large $r$, where we first employed the large $r$-approximation and
then took $\epsilon\to 0^+$ (the \lq\lq dimensional
regularization"), as can be understood from (\ref{apx}). Putting
all the above results together, we have for the potential
\begin{equation}
V=T_{D3}V_{D3}\left[1+c_7(\sin\theta_1-\sin\theta_2)^2\ln(r/\Lambda)\right].
\end{equation}
Thus we have again reproduced the logarithmic long range potential
between the $D3$ and $D7$-brane. In the language of this
subsection, the self-duality of the $B$-field, i.e $B^-=0$,
implies supersymmetry of this system. This was to be expected from
the gauge invariance of ${\mathcal{F}}=F-B$.

\sect{Minkowski  vacuum: D3/D7 bound state}

The attractive force leads the D3-brane towards D7-brane. When the
critical distance (\ref{dcritical}) is reached (where a
hypermultiplet scalar becomes massless), tachyon condensation
brings the system of these branes into a bound state: D3-brane is
dissolved on the worldvolume of the D7 brane. Supersymmetry is
restored. In this section we describe this absolute ground state,
show that it has an unbroken supersymmetry and that the D-flatness
condition is satisfied. In our previous D4/D6 model
\cite{Herdeiro:2001zb} the picture of reconfigurated branes was
suggested for the final state.  In the present case, we are able
to give a much more detailed and clear description of this state
using the idea of a D3/D7 supersymmetric bound state which has an
energy lower than that of D3 plus D7 when they are at a distance
from each other.

The Higgs stage of D3/D7 cosmological model is closely related to
the known descriptions of D0/D4 bound state in the framework of
non-commutative gauge theory \cite{Nekrasov:1998ss} as well as in
string theory and Dirac-Born-Infeld theory
\cite{Seiberg:1999vs}-\cite{Moriyama:2000eh}. The non-linear
instantons of the $\kappa$-symmetric branes in
\cite{Marino:1999af} were constructed in the context of flat
Euclidean branes. In particular, the most relevant case therein
for our application is the Euclidean D3-brane instanton. There is
no need to use the approximation of a double scaling limit, as in
\cite{Seiberg:1999vs,Marino:1999af} to describe a D3/D7 bound
state since we will employ the method to find solutions to the BPS
equation for the non-linear instanton suggested in
\cite{Moriyama:2000eh}.

The setup is a D7-brane with normal Minkowski space signature in
0,1,2,3 directions and  Euclidean signature in the
four-dimensional space parameterized by 6,7,8,9, where the
supersymmetric non-linear instanton solution for the gauge field
will be found in the presence of a constant, non-self-dual
magnetic flux $F_{mn}$  ($B_{mn}$) field. If our $F_{mn}$
($B_{mn}$) field is self-dual to start with, $F^-=0$ ($B^-=0$)
there is no Coulomb branch or inflation in our model. It would be
a limit in which all 3 FI terms vanish. So we will keep a constant
$F^-\neq 0$ ($B^-\neq 0$)  which nevertheless allows for a  Higgs
branch with unbroken supersymmetry.

\subsection{D3/D7 bound state and unbroken $\kappa$-symmetry}

In the context of the $\kappa$-symmetric D7-brane action
(\ref{D7}), we will look for a D3/D7 supersymmetric bound state
for fixed $\alpha'$.  We place our $D7$-brane in the
 background of the 10d target space, which is assumed to be flat Minkowski space with a non-vanishing  constant $ B_{mn}$ field in
$\mathbb{R}^4$ with coordinates $X^m, m=6,7,8,9$.

The equation for unbroken supersymmetry is
$\Gamma\epsilon=\epsilon $, where \cite{Bergshoeff:1997kr} \be
\Gamma=   e^{-a}\, i \sigma_2 \otimes \Gamma_{{01236789}} \ ,\ee
and \be a= {1\over 2} Y_{ik}\Gamma^{ik}\otimes \sigma^3  \ .\ee
Here the matrix $Y$ is a non-linear function of the matrix
$\calF$. In the Coulomb phase, we could choose a skew diagonal
basis for $\calF$ everywhere, since $\calF$ was constant. In
section 2 we choose $\calF_{67}=\tan \theta_1$ and
$\calF_{89}=\tan \theta_2$ leading to $Y_{67} =\theta_1$ and
$Y_{89} =\theta_2$. In the Higgs phase  $F$ is allowed to depend
on the worldvolume coordinates; therefore $\calF= F-B$ can not be
brought to a skew diagonal form everywhere. The worldvolume gauge
field $F$ is also assumed to vanish asymptotically. In such case,
the expression for the matrix $Y(\sigma)$ is a complicated
non-linear expression  in terms of $\calF(\sigma)$: \be
\exp\left\{{-{1\over 2} Y_{ik}\Gamma^{ik}\otimes
\sigma^3}\right\}= {1\over \sqrt{|\eta+\calF|}} \left[1- {1\over
2}\sigma_3\Gamma^{ik}\calF_{ik} + {1\over
8}\Gamma^{{ikml}}\calF_{ik} \calF_{ml}\right] \ . \label{Y}\ee At
large $\sigma^6, \sigma^7, \sigma^8, \sigma^9$ our matrix
$Y(\sigma)|_{\sigma \rightarrow \infty}\equiv Y^0$ depends only on
a constant matrix $B$ \be \exp\left\{{-{1\over 2}
Y^0_{ik}\gamma^{ik}\otimes \sigma^3}\right\}= {1\over
\sqrt{|\eta-B|}} \left[1+ {1\over 2}\sigma_3\gamma^{ik}B_{ik} +
{1\over 8}\gamma^{{ikml}}B_{ik} B_{ml}\right] \ .\label{Y0}\ee Now
it will be useful to introduce a difference between
$\sigma$-dependent $Y$ and its asymptotic value: \be \hat Y =
Y\left[F(\sigma)-B\right]-Y[-B] \equiv Y- Y^0= F(\sigma)+\dots \ .
\ee Here, the dots stand for terms which are non-linear in $F$ and
$B$. We can split $Y$ into a self-dual and an anti-self-dual part
$Y= Y^+ + Y^-$.

The Killing spinor equation $\Gamma \epsilon= \epsilon $ takes the
form: \bea \exp \left\{-{1\over 4} \sigma_3 \otimes \Gamma^{ik}
[Y^+_{ik} (1-\Gamma_{{6789}})+ Y^-_{ik}
(1+\Gamma_{{6789}})]\right\} i\sigma_2  \otimes
\Gamma_{{01236789}} \, \epsilon =\epsilon \ .
\label{killingInst}\eea Our spinors $\epsilon $ are constant,
therefore for $Y$ depending on $\sigma$ in an arbitrary way there
is no solution with unbroken supersymmetry. However, in contrast
to Coulomb phase, the dependence on $\sigma$ can now be
constrained in a way such that (\ref{killingInst}) has a solution.

There are two possibilities to have unbroken supersymmetry. Choose
the spinors in $\mathbb{R}^4$ to be antichiral (chiral) and
require that $Y^+$ ($Y^-$) does not depend on $\sigma$, which in
turn means that $Y^{\pm}(\sigma)- (Y^0)^{\pm}=0$,

\be (1\pm\Gamma_{{6789}})\epsilon=0 \ ,\qquad  {\partial
Y^{\pm}\over
\partial \sigma}=0 \qquad \Rightarrow  \qquad \hat Y^{\pm}(\sigma) =0
\ .\label{plus} \ee The remaining condition on spinors is \be
 \qquad \exp\left\{{-{1\over 2} \sigma_3 \otimes  \Gamma^{ik} ({Y^0})^{\pm}_{ik}}\right\} i\sigma_2  \otimes \Gamma_{{01236789}} \,
 \epsilon =\epsilon \ .
\ee

Now we have to decide which of these solutions is the correct
description of the Higgs branch of our cosmological model. We can
consider a limit in which there is no deformation due to $B$. With
antichiral (chiral) spinors in $\mathbb{R}^4$ we find the
following supersymmetric configuration: \be i\sigma_2  \otimes
\Gamma_{{0123}} \, \epsilon =\mp\epsilon \ ,\qquad  i\sigma_2
\otimes \Gamma_{{01236789}} \, \epsilon =\epsilon \ ,\qquad
F^{\pm} =0 \ .\label{barD3D7} \ee This is a $\bar {D}$3 (D3) brane
projector and a D7 brane projector, in the notation of
\cite{Bergshoeff:1997kr}.

Thus,  our solution has chiral spinors in $\mathbb{R}^4$. To have
unbroken supersymmetry, in the presence of deformation $B$,
requires that $\hat Y^-(\sigma) =0$; this has an interpretation of
a D3-brane (and not a $\bar { D3}$) dissolved into a D7-brane. It
can be shown, using (\ref{Y}) and (\ref{Y0}), that $\hat
Y^-(\sigma) =0$ is equivalent  to \be {{\cal F}_{ik}^-\over 1
+{\rm Pf} {\cal F}}= -  {B_{ik}^-\over 1 +{\rm Pf} B}\ .
\label{BPSframe}\ee Essentially the same BPS equation  for the
$\kappa$-symmetric Euclidean D3-brane was derived in
\cite{Marino:1999af}.

\subsection{ D3/D7 bound state and  deformed instantons}

In this section we shall discuss a solution of the BPS equation
(\ref{BPSframe}), that is, a slight generalization of the
nonlinear instanton solution of Seiberg and Witten
\cite{Seiberg:1999vs}, without taking their double scaling limit.
This problem was worked out by Moriyama in \cite{Moriyama:2000eh}.
Following the analysis therein,\footnote{Our case is related to
\cite{Moriyama:2000eh} by the replacements, $B \rightarrow -B$,
$\theta \rightarrow -\theta$ and $\epsilon^{ijkl} \rightarrow -
\epsilon^{ijkl}$.} we shall first show that the BPS equation
(\ref{BPSframe}) can be brought to the form
\begin{eqnarray}
\Ful^-_{ab}=\thetaul^-_{ab}\Pf\Ful \ . \label{statement}
\end{eqnarray}
We have defined  ${\cal F}^{-}_{ij}={\cal F}_{ij}-\half\epsilon_{ijkl}{\cal F}^{kl}$ and $\Pf {\cal F}={1 \over 8}\epsilon^{ijkl}{\cal F}_{ij}{\cal F}_{kl}$, while $\Ful^{-}_{ab}=\Ful_{ab}-\half\epsilon_{abcd}\Ful^{cd}$ and $\Pf \Ful={1 \over 8}\epsilon^{abcd}\Ful_{ab}\Ful_{cd}$. The indices $(i,j,\cdots)$ and $(a,b,\cdots)$ are associated respectively with the closed string metric $g_{ij}=\delta_{ij}$ and the frame metric $\delta_{ab}$. The frame metric $\delta_{ab}$ is defined with respect to the open string metric
\begin{equation}
G_{ij}=\delta_{ij}-B_{ik}\delta^{km}B_{mj} \ .
\end{equation}
The two types of indices, $(i,j,\cdots)$ and $(a,b,\cdots)$, are
converted by the vierbein $e^a_{\hspace{0.2cm}i}$ which is given
by
\begin{eqnarray}
e^a_{\hspace{0.2cm}i}=\delta^a_{\hspace{0.2cm}i}-B^a_{\hspace{0.2cm}i}
=(1-B)^a_{\hspace{0.2cm}i} \ , \qquad\mbox{with}\qquad
G_{ij}=\delta_{ab}e^a_{\hspace{0.2cm}i}e^b_{\hspace{0.2cm}j}
=(e^Te)_{ij} \ .
\end{eqnarray}
Likewise the noncommutative parameter $\theta^{ij}$ is defined by
\begin{equation}
\theta^{ij}=-\left(\frac{1}{1-B}\right)^{ik}B_{km}
             \left(\frac{1}{1+B}\right)^{mj}.
\end{equation}
On the other hand, the frame quantities $\Ful_{ab}$ and
$\thetaul_{ab}$ in (\ref{statement}) are defined as
\begin{eqnarray}
\Ful_{ab}&=&\left(\left(e^T\right)^{-1}\right)_a^{\hspace{0.2cm}i}
F_{ij}\left(e^{-1}\right)^j_{\hspace{0.2cm}b} \ ,
\label{fieldstrength}\\
\thetaul^{ab}&=&e^a_{\hspace{0.2cm}i}\theta^{ij}
\left(e^T\right)_j^{\hspace{0.2cm}b}
=-\delta^{ak}\delta^{mb}B_{km}=-B^{ab} \ . \label{theta}
\end{eqnarray}
Then the frame Pfaffian $\Pf\Ful$ is related to $\Pf F$ by
\def\det{\mbox{det}}
\begin{eqnarray}
\Pf\Ful= (\det~e)^{-1}\Pf F \ .
\end{eqnarray}

Now notice that an identity,
\begin{eqnarray}
\Ful^-=(\det~e)^{-1}\left\{(1+\Pf B)F^- -{1 \over
4}\epsilon^{klmn}F_{kl}B_{mn}B^- -\half(B^-F^--F^-B^-) \right\} \
, \label{identity}\end{eqnarray} holds, while the BPS equation
(\ref{BPSframe}) can be rewritten as
\begin{equation}
(1+\Pf B)F^-_{ij}-{1 \over 4}\epsilon^{klmn}F_{kl}B_{mn}B^-_{ij}
=-B^-_{ij}\Pf F \ . \label{BPS1}
\end{equation}
Since the BPS equation (\ref{BPSframe}) also implies that $F^-$ is
proportional to $B^-$, the third term $B^-F^--F^-B^-$ in the
identity (\ref{identity}) is vanishing. Thus we indeed find
equation (\ref{statement})
\begin{equation}
\Ful^-_{ab}=-(\det~e)^{-1}B^-_{ij}\delta^i_a\delta^j_b\Pf F
 =\thetaul^-_{ab}\Pf\Ful \ ,
\end{equation} as claimed.
\def\Abar {\underline{A}}
Now it is straightforward to solve this equation, following
Seiberg and Witten \cite{Seiberg:1999vs}. First note that eq.
(\ref{fieldstrength}) for the field strength $\underline F$ can be
restated, in terms of the frame coordinate $x^a$ and the gauge
potential $\Abar_a$, as
\begin{eqnarray}
x^a&=&e^a_{\hspace{0.2cm}i}x^i=(1-B)^a_{\hspace{0.2cm}i}x^i \ ,\\
\Abar_a&=&\left(\left(e^T\right)^{-1}\right)_a^{\hspace{0.2cm}i}A_i
=\left(\frac{1}{1+B}\right)_a^{\hspace{0.2cm}i}A_i \ .
\end{eqnarray}
Then the solution of eq. (\ref{statement}) will be given by
\begin{equation}
\Abar_a=\thetaul^-_{ab}x^bh(R) \ , \label{solution}
\end{equation}
where we have defined
\begin{equation}
R^2=\delta_{ab}x^ax^b \ .
\end{equation}
The function $h(R)$ takes the form, as in \cite{Seiberg:1999vs} and \cite{Terashima:1999tn},
\begin{equation}
h(R)=-\frac{1}{2\Pf\thetaul^-}
\left(1-\sqrt{1+\frac{4C\Pf\thetaul^-}{R^4}}\right) \ ,
\end{equation}
where the constant $C$ must be positive but otherwise arbitrary,
as $\Pf\thetaul^-$ is negative. Below, we will take $C$ to be
$-8N$, with $N$ being integer.

In the presence of the noncommutative parameter $\thetaul^-$ or
equivalently the $B^-$ field, the singularity of $U(1)$ instanton
is improved, as discussed in \cite{Seiberg:1999vs}. In particular
the solution (\ref{solution}) has a finite instanton number,
though the gauge potential itself is singular as we go to the UV:
\begin{eqnarray}
{1\over 16\pi^2}\int_{\mathbb{R}^4}  F\wedge  F ={1 \over
8\pi^2}\int d^4x\Pf\Ful =-{1 \over
8}\Pf\thetaul^-\lim_{R\rightarrow 0} \left(R^4h^2\right) =N \ .
\end{eqnarray}
Note that the final result does not depend on the value of
$\thetaul^-$, but it was crucial to keep $\thetaul^-$ finite in
the procedure of getting finite instanton number. Had we taken the
limit $\thetaul^- \rightarrow 0$ first, the instanton number would
have been divergent, as $h^2(R)$ grows as $R^{-8}$, as $R$ goes to
zero, which corresponds to the fact that there is no Abelian
instanton with finite energy.

In our cosmological scenario, the nonvanishing $\thetaul^-$ is the
seed of the inflation, giving positive vacuum energy and producing
an instability. In particular our model realizes the slow-roll
inflation, as discussed in the previous section. After the
slow-roll inflationary stage, the universe eventually goes through
the tachyon condensation or `waterfall stage', and ends up with
the Minkowski vacuum. As we mentioned earlier, this endpoint
vacuum is described by a non-marginal bound state of D3 and
D7-branes that corresponds to the Higgs phase of the gauge theory
of the D3/D7 system with the FI term provided by the nonvanishing
$\thetaul^-$. It is quite well-known that the D3-branes on the
D7-branes can be thought of as instantons on the D7-branes due to
the Chern-Simons coupling,
\be S_{CS}= {1\over 16 \pi^2}
\int_{D7}C_{4}\wedge F \wedge F= N \int_{D3}C_{4}.
\ee
In the above we have presented a D7-brane probe approximation of the Higgs phase where the D3-brane is described by a nonlinear or deformed instanton on the $U(1)$ DBI theory on the D7-brane.

However, given the fact that the Higgs phase of this D3/D7 system is actually identical to the noncommutative generalization of the ADHM construction of instantons, one may make a step further and argue that our Minkowski vacuum is described by the noncommutative instanton of Nekrasov and Schwarz \cite{Nekrasov:1998ss}, improving the nonlinear instanton solution that we have studied above. The noncommutative instanton solution is non-singular even at $R=0$, resolving the singularity of the zero-size instantons, and the action is of course finite.

It is to be stressed that we have thus found a possible connection of the cosmological `constant' in spacetime ($0123$-space) and the noncommutativity in internal space ($6789$-space):
\begin{eqnarray}
\Lambda\,
(\mbox{spacetime cosmological constant})
\quad \Leftrightarrow \quad
\thetaul^-\,
(\mbox{noncommutativity in internal space}).
\nonumber
\end{eqnarray}
We have generated the cosmological constant by turning on $B^-$ or
equivalently $\thetaul^-$, which can eventually be interpreted as
the noncommutative parameter in the internal space at the end of
our cosmological evolution, the Higgs phase, of our D3/D7 model.
Also it may be somewhat suggestive that the non-vanishing
non-commutativity parameter, i.e. the cosmological constant,
serves, in a way, to resolve the singularity at the end of
inflation.

\sect{A compactified type IIB setup}

In section 2 we studied a system of separated
 D3/D7 with world-volume
two-form fluxes such that there is no supersymmetry. This was the
Coulomb phase of our model. Because of spontaneously broken
supersymmetry the $D3$-brane would be attracted towards the $D7$
and eventually it would fall into the $D7$ as an instanton. Due to
the presence of fluxes (as FI term) the instanton $-$ which is the
dissolved $D3$ $-$ will behave as a non-commutative instanton. The
point like singularity of the instanton is therefore resolved.
This is the Higgs phase studied in section 3 and now it is
supersymmetric. Given such a setup, there are two interesting
questions that we can ask at this stage:

\noindent (i) Is it possible to get a {\it compactified}
four-dimensional model for this setup? We would then need to
compactify our type IIB model on a six-dimensional base. But this
raises the problem of anomalies or equivalently charge
conservation. Therefore what we need is a configuration in four
dimensions which is anomaly free.

\noindent (ii) Is it possible to study the metric for the system
for both the Coulomb and the Higgs phase? In the Coulomb phase  we
expect the metric to have an inherent time dependence. In the
Higgs phase $-$ now because of supersymmetry $-$ we expect a time
independent metric.

In this section we shall see that both these questions can be
partially answered from a type IIB setup involving both branes and
orientifold planes. In section 5 we describe its M-theory lift.

\subsection{The model}

To get a model in four dimensions we have to compactify our
$D3/D7$ system on a six dimensional space. The branes are oriented
as in Table 1. Therefore we need to compactify directions $x^{4,
5, 6, 7, 8, 9}$, which raises the question of charge conservation.
Both the $D3$ and the $D7$ branes are {\it points} on the compact
two space $x^{4, 5}$. Therefore to cancel the charges of the
branes we have to put {\it negatively} charged objects in our
model. These have to be orientifold planes, since anti-branes
would annihilate with the branes destroying the dynamics. However
in string theory an orientifold operation is an elaborate
mechanism which requires various other branes and planes to get a
completely consistent picture. We now describe one consistent
setup which satisfies all the requirements of a compact model.

The model that we take to get a compactified type IIB setup has
been studied earlier in a different context for supersymmetric
cases. This involves compactifying type IIB on $K3 \times
T^2/{\mathbb Z}_2$ where the ${\mathbb Z}_2$ is an orientifold
operation:
\begin{equation}
{\mathbb Z}_2 = \Omega \cdot (-1)^{F_L} \cdot {\mathcal I}_{45} \
.
\end{equation}
The torus $T^2$ is oriented along $x^4, x^5$; ${\mathcal I}_{45}$
is the orbifold projection with the action $x^{4,5} \to -x^{4,5}$;
$\Omega$ is the worldsheet orientation reversal and $(-1)^{F_L}$
changes the sign of the left moving space-time fermions. The $K3$
is oriented along $x^{6,7,8,9}$. Under the $\Omega \cdot
(-1)^{F_L}$ operation the various fields of type IIB change as:

\begin{center}
\begin{tabular}{*{9}{|c}|}
\hline
 $\mbox{}$ & $\Omega$ & $(-1)^{F_L}$ & $\Omega \cdot(-1)^{F_L}$  \\
\hline
$g_{MN}$ & + & + & +  \\
\hline
$B^{NS}_{MN}$ & $-$ & + & $-$ \\
\hline
$B^{RR}_{MN}$ & + & $-$ & $-$ \\
\hline
 $\varphi$ & + & + & + \\
\hline $\tilde{\varphi}$ & $-$ & $-$ & + \\
\hline
$D^{+}$ & $-$ & $-$ & + \\
\hline
 \end{tabular}
\end{center}
\begin{center}
Table 3 \end{center} Taking further into account ${\mathcal
I}_{45}$, we see from the above table that $B$ fields along
$x^{6,7,8,9}$ directions are projected out. The only $B$ fields
that survive must have one leg along $x^{4,5}$. These appear as
gauge fields in four-dimensions or as scalars. Other fields which
have one leg either along $x^4$ or $x^5$ are also removed from the
spectrum. For example if we denote
\begin{equation}
m,n = x^{4,5}, ~~~ \alpha, \beta = x^{6,7,8,9}, ~~~~ \mu, \nu =
x^{0,1,2,3} \ ,
\end{equation}
then reducing over $T^2/{\mathbb Z}_2$ we have, in addition to a
metric $g_{\mu \nu}$, the following spectrum of scalars and
vectors from the supergravity fields (in terms of four dimensional
multiplets):
\begin{center}
 \begin{tabular}{*{9}{|c}|}
 \hline
 $\mbox{}$ & Scalars & Vectors \\
\hline
$g_{MN}$ & $g_{mn}, g_{\alpha \beta}$ & $g_{\alpha \mu}$ \\
\hline
$B_{MN}^{NS, RR}$ &$B_{m \alpha}^{NS, RR}$  &$B_{m \mu}^{NS, RR}$ \\
\hline
$\varphi, {\tilde \varphi}$ & $\varphi, {\tilde \varphi}$  & $-$ \\
\hline
$D^{+}_{MNPQ}$ &$D^{+}_{mn\alpha \beta}, D^{+}_{0123}, D^{+}_{\mu\nu\alpha
\beta}$  & $D^{+}_{\alpha \beta \gamma \mu}$ \\ \hline
 \end{tabular}
\end{center}
\begin{center}
Table 4
 \end{center} However $K3$ further cuts down the spectrum as
the first Betti number is zero (so that no one forms can exist).
In particular it kills $g_{\alpha \mu}, B^{NS,RR}_{m\alpha}$ and
$D^+_{\alpha \beta \gamma \mu}$. It also restricts the number of
two forms on it to be
\begin{equation}
b_2 = b_{11} + 2 b_{20} = 22 \ ,
\end{equation}
corresponding to the number of four dimensional scalars from the
metric. This is the supersymmetric spectrum of the model. This
spectrum is, as we know, incomplete. The complete spectrum
involves additional $16~ D7$-branes as $4~ O7$-planes. The
$O7$-planes are located at the four orbifold singularities of
$T^2/{\mathcal I}_{45}$. To cancel charges locally we need $4~D7$
branes on top of each orientifold planes. Therefore we have 7
branes wrapping the $K3$. This is a familiar model in the F-theory
context. As discussed in  \cite{sen96} the axion-dilaton is
constant and could be arbitrary.\footnote{There are however some
special points in the constant coupling moduli space where the
seven branes are distributed in some specific way which gives rise
to exceptional global symmetry on the four-dimensional gauge
theory \cite{dasmukhi}. These points are at strong coupling. We
ignore these subtleties because they will not affect the result.}

\subsubsection{The orbifold limit}

Naively from the supergravity point of view it is difficult to
argue the existence of the tetrahedron $T^2/{\mathcal I}_{45}$
since the total {\it source} is zero at the orbifold point.
However, perturbative string theory `feels' it. Thus, a D3-brane
feels the orientifold background, even if both charge and tension
are cancelled by the presence of both orientifolds and
D-branes.\footnote{At this point putting a probe looks arbitrary
and it seems like we have not cancelled its charge on the compact
space. In the next section we shall argue the consistency of this
from the M-theory point of view.}
 Let us concentrate on one bunch of $4~D7 + O7$. The gauge theory
on the $D3$ is a $Sp(1) \equiv SU(2)$ Seiberg-Witten theory with 8
supercharges whose matter hypermultiplets are given by the D3/D7
strings. The strings which go around the $O7$-plane and come back
give the massive $W^{\pm}$ bosons charged under the $U(1)$ gauge
multiplets (the string starting from and ending on
 the $D3$).

The location of the $D3$-brane in the $x^{4,5}$ corresponds to the
expectation value of the complex chiral field in the adjoint of
$SU(2)$ $-$ this is  the {\it inflaton} field in our model $-$
which belongs to the vector multiplet. Denoting this as $\Phi =
x^4+ix^5$ we can diagonalise this
 to
\begin{equation}
\Phi=
 \left(
 \begin{array}{cccc}
d & 0 \\
0 & -d
 \end{array}
 \right).
\end{equation}
The classical massless point is $d=0$ and the $W^{\pm}$ have masses $2d$.

There are also strings which start from and end on the
$D7$-branes. They give rise to the $SO(8)$ gauge fields on the
$D7$-branes. These strings survive the orientifold projection
because the vertex operators are neutral under
$\Omega$-projections. However there could be strings which start
from a $D7$, cross an $O7$ and come back to the $D7$. For these
strings $\Omega$-projections remove the first massless vector
multiplets charged under $SO(8)$. The surviving states are the
next stringy oscillations which are massive and non-BPS
\cite{sennonbps}.

The {\it Coulomb} branch of our model is when we switch on  gauge
flux ${\cal F}$ on the seven branes. As discussed earlier this
flux is {\it non-primitive} i.e non-self dual. This breaks
supersymmetry spontaneously and therefore the $D3$ probe would be
attracted towards the D7. This motion creates an instability in
the system which triggers off non-perturbative corrections on the
background.

\subsubsection{Away from orbifold limit}

Under non-perturbative corrections each $O7$-plane splits into two
$(p,q)$ seven branes (where $(p,q)$ seven branes are defined as
branes on which $(p,q)$ strings can end $-$ $p,q$ denoting the two
background values of $B$ fields). In a theory with 16 bulk
supercharges this is explicitly demonstrated in \cite{sen96,bds}.
We expect something similar for our case too. However our main
concern is to study an {\it isolated} and compactified D3/D7
system. This can be achieved by pulling a $D7$ from the bunch and
study its dynamics in the presence of a probe $D3$-brane. This way
we can isolate the other $O7-D7$ dynamics from our inflationary
model.

Our last concern is the anomaly cancellation. For the SUSY setup
of our model (with self-dual ${\cal F}$) the argument of anomaly
cancellation is simple. Let us make two T-dualities along
directions $x^{4,5}$ of the torus $T^2$. Under this
\begin{equation}
\Omega \cdot (-1)^{F_L} \cdot {\mathcal I}_{45}~~ \to ~~ \Omega
\end{equation}
This is now type I theory on a torus which is further dual to
heterotic string on a torus. As we know all anomalies $-$ gauge
and gravity $-$ are cancelled for this theory and therefore we
expect the same for type IIB background.

In our case however the background is non-supersymmetric and hence
T-dualities are subtle. So anomaly cancellation has to be checked by
explicitly doing the one loop graph for the given background. However as
we shall discuss in the ensuing section, anomaly cancellation can be
equated in this case to charge cancellation on a compact manifold in M-theory.
For this case therefore we can argue that anomalies do get cancelled.

Before we end this section let us summarise the situation. First, the full
matter content of our theory:

\noindent (1) {\it Gauge fields on D3-brane}: These are the
$SU(2)$ gauge fields broken to $U(1)$ at any generic point of our
model. In fact there is never an enhancement to $SU(2)$ in this
setup, so it remains $U(1)$ all through
\cite{Seiberg:1994rs,Seiberg:1994aj,sen96,bds}. If we take large
number of D3-branes then the situation can be more subtle.

\noindent (2) {\it Gauge fields on the D7-branes}: These are the
$SO(8)$ gauge fields on the seven branes. In our case we have
switched on a background $U(1)$ field on one of the D7 which
breaks supersymmetry in our model. This field, ${\cal F}$, acts as
an FI term in the D3 worldvolume theory.

\noindent (3) {\it Massive hypermultiplets}: These are generated
by $3-7$ strings which survive the $\Omega$-projections. In
general they are massive.

\noindent (4) {\it Bulk modes}: These come from the bulk
supergravity modes. Under $\Omega$-projection many of the bulk
fields are removed from the spectrum as discussed above. The
surviving ones undergo further projection due to the $K3$
manifold. In the end we get a metric, scalars and vectors in four
dimensions. These vectors are in addition to the $SO(8)$ and the
$U(1)$ vectors of the D7 and D3 respectively.

\noindent (5) {\it Stable non-BPS states}: These are actually the
modes that survive $\Omega$-projection on the D7-branes coming
from the 7-7 strings that go around the O-plane. They are all
massive and receive sizable quantum correction.

\noindent (6) {\it Modes on $O$-planes}: They are non existent
when we neglect non-perturbative corrections. However in our model
we expect the corrections to be sizable as we have isolated a D7
from the rest of the dynamics. Therefore the $O$-planes no longer
remain non-dynamical. Now $(p,q)$ strings can end on them and
therefore they support matter fields.

\noindent (7) {\it Junctions and networks}: Additional
states can also arise in our
 setup. A D7 can be connected to two different $(p,q)$ seven branes via a
junction or a string network. They contribute additional states. But these are
in general massive, so we can neglect them.

{}From the above analysis we see that most of the other matter
fields coming from branes and planes are actually massive, and in
the limit when we separate one D7 from the whole crowd, we can in
principle study a compactified theory with few massless matter
fields (including gravity). This is what we have tried to achieve
here.

\section{M-theory on a four-fold with G-fluxes}

The above setup can be further simplified by lifting it to
M-theory. All the complicacies of branes and planes now disappear
in M-theory. The system of $16 D7 + 4 O7$ located at four points
of $T^2/{\mathcal I}_{45}$ in IIB simply becomes a $T^4/{\mathbb
Z}_2$ orbifold of $K3$
 in M-theory. In other words the whole setup in IIB is
actually just M-theory compactified on $T^4/{\mathbb Z}_2 \times
K3$! Denoting the $T^4$ directions as $x^{3,10,4,5}$ the ${\mathbb
Z}_2$ operation sends (notice this is distinct from the ${\mathbb
Z}_2$ of section 4):
\begin{equation}
 x^{3,10,4,5} ~~~ \stackrel{{\mathbb Z}_2}\longrightarrow ~~~ -
 x^{3,10,4,5}\ .
\end{equation}
The original positions of the branes and planes become orbifold
singularities in M-theory. The non-supersymmetric flux on one of
the seven branes in type IIB becomes localized G-flux near one of
the orbifold singularities. Supersymmetry is broken by choosing a
non-primitive G-flux. Before we go into the details of the
M-theory setup let us summarise the situation in the following
table:
\begin{center}
 \begin{tabular}{*{9}{|c}|}
 \hline
 $\mbox {\bf Type~IIB}$ & ${\bf M-theory}$   \\
\hline $T^2/ (\Omega \cdot (-1)^{F_L} \cdot {\mathbb Z}_2) \times
K3$ & $T^4/{\mathbb Z}_2
\times K3$ \\
\hline
4($O7 + 4~D7$) & 4~orbifold~fixed~points \\
\hline
$D3$ & $M2$ \\
 \hline
${\cal F}$~on~$D7$ & Localised~G-flux~at~fixed-points \\
\hline
Coulomb phase & Non-primitive~G-flux \\
\hline
Higgs phase & Primitive~G-flux \\
\hline Away~from~orientifold~limit & $T^4/{\mathbb Z}_2 \ \ \rightarrow $ \ \ Smooth~$K3$ \\
\hline
 \end{tabular}
\end{center}
\begin{center}
Table 5 \end{center} The model that we are going to use is
M-theory compactified on a four-fold with G-fluxes switched on. To
get a ${\cal N} = 2$ theory we have to compactify M-theory on
$K3\times K3$. We shall take one of the $K3$ to be a torus
fibration over a ${\mathbb C}P^1$ base (see figure 2). The
Weierstrass equation governing the background is given by
\begin{equation}
y^2 = x^3 + x f(z)  + g(z) \ ,
\end{equation}
where $x,y,z \in {\mathbb C} P^1$, $f(z)$ is a polynomial of
degree eight, and $g(z)$ is a polynomial of degree 12 in $z$. As a
concrete example let us consider \cite{sen96}
\begin{equation}
g(z)= \prod^4_{i=1} (z-z_i)^3,~~~~f(z)= a \prod^4_{i=1} (z-z_i)^2
\ , \label{orbik3}
\end{equation} where $a$ is a constant. This basically describes a
torus at every point of ${\mathbb C}P^1$ labelled by the
coordinate $z$. The modular parameter $\tau(z)$ of the torus is
determined in terms of $f^3/g^2$ through the relation
\begin{equation}
j(\tau)={(\theta_1^8(\tau)+\theta^8_2(\tau)+\theta^8_3(\tau))^3\over
\eta(\tau)^{24}} = {4.(24f)^3\over 27g^2+4f^3}= {55296a^3 \over
27+4a^3} \ .
\end{equation}
\begin{figure}[h!]
\begin{picture}(75,0)(0,0)
\put(160,140){$T^2$} \put(170,90){${\mathbb C}P^1$} \put(235,
115){K3} \put (60,140){$(x^3,x^{10})$} \put(50,90){$(x^4,x^5)$}
\put(0,25){$(x^6,x^7,x^8,x^9)$} \put(190,25){K3}
\end{picture}
\centering \epsfysize=13cm
\includegraphics[scale=0.6]{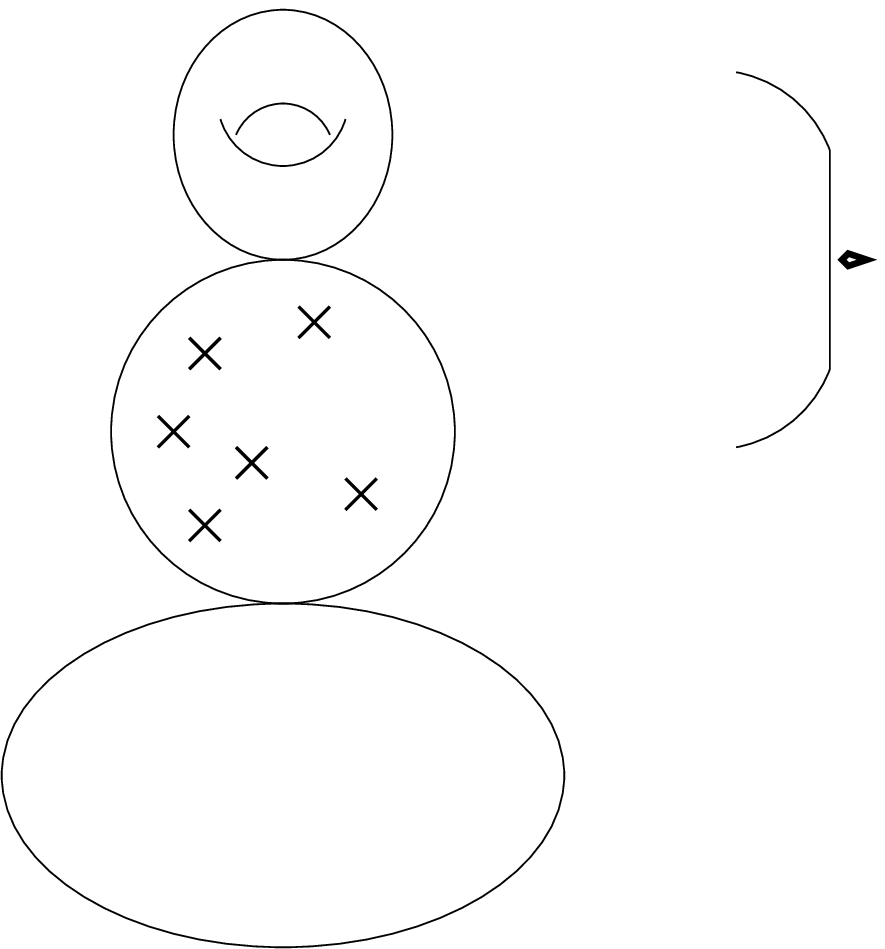}
\caption{The ``snowman" fibration of the $K3\times K3$ four-fold.
The crosses indicate points on the ${\mathbb C}P^1$ basis at which
the fibre tori degenerate. In the orbifold limit of the `top' K3
there will be four such points.}
\end{figure}
The number of points at which the torus degenerates is given in
terms of the zeroes of the polynomial
\begin{equation}
\Delta\equiv 4f^3+27g^2 \ .
\end{equation}
In terms of our choice of $f$ and $g$ the number of points at
which the torus degenerates is given as:
\begin{equation}
\Delta=(4a^3+27) \prod^4_{i=1} (z-z_i)^6 = 0 \ .
\end{equation}

To go to type IIB theory we have to shrink the fiber torus to zero
size. Let us study this decomposition carefully. We first denote
the base $K3$ by coordinates $x^{6,7,8,9}$ and the other $K3$ by
$x^{4,5,3,10}$ in M-theory. The spacetime is $x^{0,1,2}$. In type
IIB therefore we get $16 D7$ and $4 O7$ planes located at points
on a ${\mathbb C}P^1$ along directions $x^{4,5}$. These branes and
planes wrap the other $K3$ along $x^{6,..,9}$. The spacetime is
now {\it four} dimensional: $x^{0,1,2,3}$. From our choice of
$f,g$ the $K3$ is elliptically fibered and therefore the ${\mathbb
C}P^1$ base in IIB is given by $T^2/ (-1)^{F_L} \cdot \Omega \cdot
{\mathcal I}_{45}$
 where $\Omega$ is the orientifold operation,
$(-1)^{F_L}$ changes the sign of the left moving fermions and
${\mathcal I}_{45}: x^{4,5}\to -x^{4,5}$.

To verify the above analysis one can make a further T-duality
along, say, $x^3$. Using the fact that \cite{dabholkar}
\begin{equation}
\Omega ~\rm{in} ~IIB~~~ \stackrel{T_3}\longrightarrow ~~~
{\mathcal I}_3 \Omega ~\rm{in}~IIA
\end{equation}
we can show that IIA theory is now on
\begin{equation}
{T^3\over (-1)^{F_L} \cdot \Omega \cdot {\mathcal I}_{345}}
\end{equation}
This however is dual to M-theory on $K3$ \cite{sen97} thus proving
our result.

Equation (\ref{orbik3}) describes the orbifold limit of K3. We
assume this is the initial stage of our dynamical process. In
terms of type IIB language, we place a D3-brane at the center of
mass of the $4\times(4D7/O7)$ setup on one side of the ``pillow"
$T^2/{\mathbb Z}_2$ and another on its diametrically opposite
side, as in figure 3. We turn on the same gauge fluxes on all of
the four fixed points. The logarithmic potential of section 2
creates a force between each pair of D3 and D7 branes, if we
assume the size of the $T^2/{\mathbb Z}_2$ to be large enough.
However in this configuration the total force between D3 and D7
branes is balanced, and the logarithmic potential being
approximately flat
 leads to a nearly de Sitter evolution. Quantum fluctuations will
destabilize the system allowing both the D3-brane to move towards
some D7-brane and the D7-branes to move away from the orbifold
fixed points. This is the beginning of the Coulomb phase or
equivalently the inflationary stage (figure 4). As the system
starts evolving we do not expect the coupling to remain constant
anymore. The whole system moves away from the orientifold limit,
which in M-theory language means we have a generic K3. Finally, we
expect the D3 to fall into one particular D7-brane as a
non-commutative instanton as explained in section 3. This is the
supersymmetric configuration that has been studied in M-theory and
which we now describe.

\begin{figure}[h!]
\begin{picture}(75,0)(0,0)
\end{picture}
\centering \epsfysize=13cm
\includegraphics[scale=0.6]{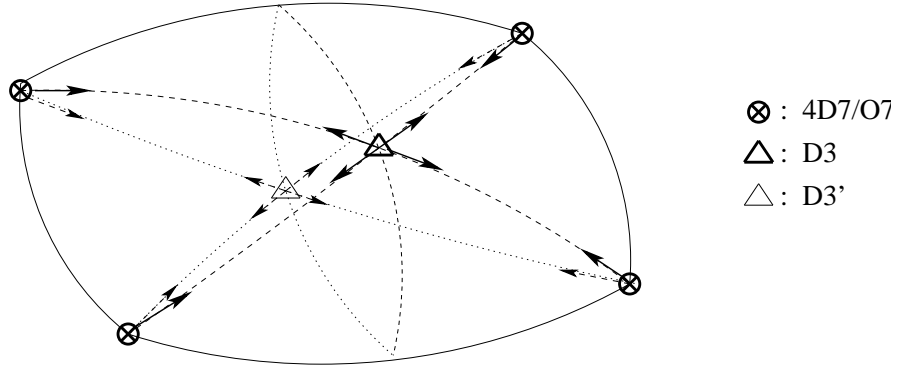}
\caption{The initial brane configuration on the ``pillow"
$T^2/{\mathbb Z}_2$.}
\end{figure}

\begin{center}
\begin{figure}[h!]
\begin{picture}(75,0)(0,0)
\end{picture}
\centering \epsfysize=13cm
\includegraphics[scale=0.6]{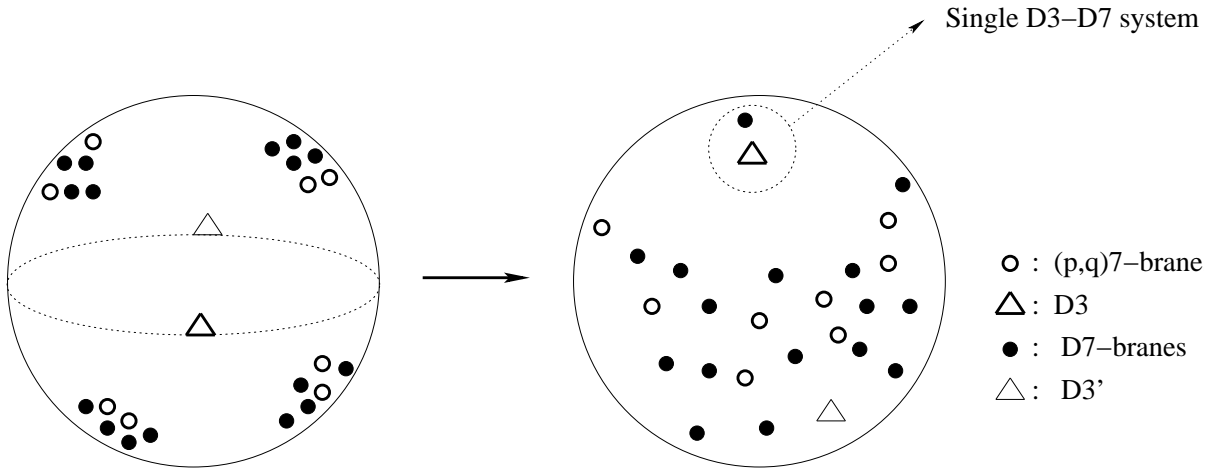}
\caption{The Coulomb branch. As the system is driven away from the
unstable point of figure 2, $T^2/{\mathbb Z}_2  \ \rightarrow \
{\mathbb C}P^1$, the orientifold planes split into $(p,q)$
7-branes, and the D3-brane will eventually fall into one D7-brane
as an instanton.}
\end{figure}
\end{center}
\subsection{Supersymmetric setup: Higgs phase}
Let us first summarise the known properties of the fourfold that
allows for a supersymmetric configuration:

$\bullet$ The 4-fold vacua has a tadpole anomaly given by $\chi /24$
where
$\chi$ is the Euler characteristics of the 4-fold.  If $\chi /24$ is
integral, then the anomaly can be cancelled by placing a sufficient
number
of spacetime filling M2-branes $n$
 on points of the compactification manifold. There
is also another way of cancelling the anomaly and this is through
G-flux. The G-flux contributes a $C$ tadpole through the
Chern-Simons coupling $\int C\wedge G \wedge G$. When  $\chi /24$
is not integral then we need both the branes and the G-flux to
cancel the anomaly. The anomaly cancellation formula becomes
\begin{equation}
{\chi \over 24} = {1\over 8\pi^2}\int G \wedge G + n \ ,
\end{equation}
which must be satisfied for type IIA or M-theory.

$\bullet$ If we denote the spacetime coordinates by $x^{\mu}$ where
$\mu
= 0,1,2$ and the internal space by the complex coordinates $y^a, a = 1,..,
4$
then in the presence of G-flux the metric becomes a warped one
\begin{equation}
ds^2 = e^{-\phi(y)}\eta_{\mu\nu}dx^{\mu}dx^{\nu}+ e^{ \phi(y)/2}
g_{a{\bar b}}dy^ady^{\bar b} \ ,
\end{equation}
with the G-flux satisfying the condition
\begin{equation}
G = *G, ~~~~~ J \wedge G = 0 \ ,
\end{equation}
where the Hodge star acts on the internal 4-fold with metric $g$
and $J$ is the Kahler form of the 4-fold. There is also another
non-vanishing G given in terms of the warp factor as
$G_{\mu\nu\rho a} = \epsilon_{\mu\nu\rho} \del_a e^{-3\phi/2}$.
The warp factor satisfies the equation
\begin{equation}
 \label{wareq}
\Delta e^{3\phi/ 2}= *\left[4\pi^2 X_8 - {1\over 2} G \wedge G -
4\pi^2 \sum^n_{i =1} \delta^8(y - y_i)\right] \ ,
\end{equation}
where the Laplacian and the Hodge * is defined with respect to
$g$, and $X_8$ is an eight-form constructed out from curvature
tensors given in (\ref{x8}).

 The G-flux can now decompose in two ways to give different {\it
supersymmetric} backgrounds in type IIB. Let us choose the
following basis of one-forms for the fiber torus (with modular
parameter $\tau$) \cite{drs}.
\begin{equation}
dz = dx + \tau dy \ , ~~~~~~d{\bar z}= dx + {\bar \tau} dy \ .
\end{equation}
In terms of spacetime coordinates the directions $y,x$ are $x^3,
x^{10}$. The ${\mathbb CP}^1$ base has coordinates $x^4,x^5$ and
the other $K3$ is oriented along $x^{6,7,8,9}$. The ``flat''
directions are $x^{0,1,2,3}$.

Decomposing G-flux as
\begin{equation}
{G\over 2\pi} = dz \wedge \omega - d{\bar z} \wedge *\omega \ ,
\end{equation}
where $\omega$ has one leg along ${\mathbb C}P^1$ and two along
K3, we get in type IIB the following three forms:
\begin{equation}
H_{NS}= \omega - *\omega, ~~~~~ H_{RR}= \omega\tau - *\omega {\bar
\tau} \ .
\end{equation}
The anomaly cancellation condition will now become, in type
IIB
theory
\begin{equation}
{\chi \over 24}= n + \int H_{RR} \wedge H_{NSNS} \ ,
\end{equation}
where $n$ is the number of D3 branes.

We can also decompose
G-fluxes as:
\begin{equation}
{G\over 2\pi} = \sum_{i=1}^k F_i \wedge \Omega_i \ ,
\end{equation}
where $i= 1,2..k$ are the number of points at which the fiber
torus degenerates, and $\Omega_i$ are the normalisable harmonic
forms localized at the singularities. The physical significance of
$F_i$ are the {\it gauge} fields that would appear on the 7-branes
when we go to type IIB by shrinking the fiber torus. The fact that
locally a $K3$ looks like a Taub-NUT space implies that there
exist a harmonic two form which is anti-self-dual and
normalisable. It is not difficult to identify the harmonic two
forms with $\Omega_i$. The anomaly cancellation condition or
equivalently the warp factor equation will now become in type IIB
theory:
\begin{equation}
\label{warpfact} \Delta e^{3\phi/2} = \ast_{\cal B} \sum_{i=1}^k
[F_i\wedge F_i + tr(R \wedge R) ]\delta^2(z-z_i) \ ,
\end{equation}
where $z_i$ are the positions of the seven branes.

Hence the general G-flux background will be
\begin{equation}
{G\over 2\pi} = \sum_i F_i \wedge \Omega_i + dz \wedge { \omega_1}
+ d{\bar z} \wedge { \omega_2} \ ,
\end{equation}
where $i$ denotes the number of singularities or equivalently
number of seven branes (in IIB).

The self-duality of $G$ flux does not imply $F$ to be self dual.
This is the precise condition for the existence of non-commutative
instanton.

Since the background is supersymmetric we now expect the metric to
be given by the analysis of \cite{bbsis1,bbsis2}. The metric has
the form of a $D3$ brane metric and the cosmological constant
vanishes.

\subsection{Coulomb Phase}

As discussed in \cite{bbsis2} a $(2,2)$ G-flux background is
non-supersymmetric if the G-flux does not satisfy the primitivity
condition i.e we require
\begin{equation}
g^{a{\bar b}}G_{a{\bar b}c{\bar d}} \ne 0 \ .
\end{equation}
Therefore let us choose a generic background for the G-flux as:
\begin{equation}
{G\over 2\pi} = dz \wedge \tilde{\omega}_1 + d{\bar z} \wedge
\tilde{\omega}_2 \ .
\end{equation}
We do not impose any conditions on $\tilde{\omega}_{1,2}$ as
$g^{a{\bar b}}G_{a{\bar b}c{\bar d}} \ne 0$. The number $n$ of
D3-branes can be tuned by choosing some appropriate values for
$\tilde{\omega}_{1,2}$ satisfying
\begin{equation}
n = 24 - \int (\tilde{\omega}_1 + \tilde{\omega}_2) \wedge (\tau
\tilde{\omega}_1 - {\bar \tau} \tilde{\omega}_2) \ .
\end{equation}
Since the system is {\it not} supersymmetric we expect the D3 to be moving
towards the seven branes. And therefore the metric ansatz cannot be a
static one as in \cite{bbsis1,bbsis2}.

\subsection{Higher order corrections}

Before we end this section let us make some remarks on higher
order corrections in eleven dimensional supergravity which were
briefly alluded to in the earlier sections. In our model there are
two different sources of higher order corrections
\cite{Tseytlin:2000sf}:

\noindent(1) {\it Curvature corrections}: In the M-theory metric the
higher curvature corrections come from the
following terms in the eleven dimensional lagrangian:
\begin{equation}
\int d^{11}x \sqrt{-g} {\Big [}J_0 - {1\over 2}E_8 {\Big ]} - \int
C \wedge X_8 \ .
\end{equation}
The various terms appearing in the above equation are defined as:
\begin{equation}
J_0 = 3 \cdot 2^8 {\Big (} R^{MNPQ}R_{RNPS}R_M^{~~TUR}R^S_{~~TUQ}
+ {1\over 2} R^{MNPQ}R_{RSPQ}R_{M}^{~~TUR}R^S_{~~TUN} {\Big )} +
{\mathcal O}(R_{MN})  \ ,
\end{equation}
\begin{equation}
E_8 = {1\over 3!} \epsilon^{ABCN_1...N_8}\epsilon_{ABCM_1...M_8}
R^{M_1M_2}_{~~~~~~N_1N_2} ... R^{M_7M_8}_{~~~~~~N_7N_8} \ ,
\end{equation}
\begin{equation}
X_8 = {1\over 3 \cdot 2^{10} \cdot \pi^4} {\Big [}
\rm{tr}~{\mathcal R}^4 - {1\over 4} (\rm{tr}~{\mathcal R}^2)^2
{\Big ]} \ . \label{x8}
\end{equation}
Now assuming that our warp factor is everywhere smooth, then integrating out
(\ref{wareq}) we have the required anomaly cancellation equation for the
four-fold. However if we put $X_8 = 0$ then observe that there could be
{\it no} non-trivial background fluxes in the model. The
only consistent solution
then is to take $G = 0$ and $n = 0$. This is the key issue which we think
is important to have the required background.

Another point which follows immediately is that since $J_0$ and
$E_8$ are of the same form (and order) as $X_8$ it will be
inconsistent to neglect them and consider only $X_8$. Therefore to
have a consistent picture in the presence of branes and planes we
{\it cannot} neglect higher order curvature terms. Also, this is a
possible way to go around the no-go theorems of Gibbons
\cite{Gibbons:85} and Maldacena and Nu\~nez \cite{Maldacena:01}.

\noindent (2) {\it G-flux corrections}: In our model G-flux could
also contribute higher order corrections. These corrections have
not been worked out in any detail. However they are believed to be
of the general form
\begin{equation}
\sum_{m,n,p}~ \alpha_{mnp} \sqrt{-g}~ (\del G)^m G^n R^p
\label{alpha} \ ,
\end{equation}
where the coefficients $\alpha_{mnp}$ are not known in
general\footnote{For some special cases when $n = p =0$ and $m =
4$ they are calculated in \cite{Deser:2000xz}.}. As long as we
take the background $G$ fluxes to be smooth and small we can in
principle neglect these correction. For sharply varying G-fluxes,
however, there could be sizable corrections which may destabilize
the background.

\section{Discussion}

In our paper the main emphasis  was on the derivation of the gauge
theory with the particular potential and FI terms from type IIB
string theory. When this gauge theory is coupled to gravity in
four dimension it  leads to hybrid inflation. As briefly discussed
in sections 4 and 5 we expect the configuration of D7-branes  and
O7-planes in the presence of  D3-branes (at the equilibrium points
on the ``pillow'') will lead to a $dS_4$ space as the beginning of
inflation. Quantum fluctuation will trigger off inflation in this
system.

The Coulomb phase of the cosmological D3/D7 setup describes a
period of slow- roll inflation in de Sitter valley and in the
Higgs phase describes a supersymmetric ground state with vanishing
cosmological constant. As discussed in the text, there is a
compactified picture in M-theory on $K3 \times K3$ manifold with a
choice of G-flux on it. When the flux is non-primitive the
background is non-supersymmetric and when the background is
primitive, supersymmetry is restored. We believe the various
moduli in this setup may be fixed by the choice of G-flux and
higher derivative corrections described in section 5
\cite{Gukov:1999ya,bbsis2}. We will leave this problem for future
work.

\section *{Acknowledgments}
We benefited from discussions with K. Becker, D. Freedman, G.
Gibbons, S. Kachru, N. Kaloper, L. Kofman, A. Linde, A. Peet, M.
M. Sheikh-Jabbari, A. Sen, L. Susskind and A. Tseytlin. This work
is supported by NSF grant PHY-9870115. The work of K.D. is
supported in part by David and Lucile Packard Foundation
Fellowship 2000 $-$ 13856. The work of C.H. is supported by grant
SFRH/BPD/5544/2001 (Portugal). The work of S.H. is supported by
the Japan Society for the Promotion of Science.


\end{document}